\documentclass[pre,a4paper,onecolumn,showpacs,showkeys,nofootinbib,floatfix]{revtex4}
\usepackage{amsfonts}
\usepackage{amsmath}
\usepackage{amssymb}
\usepackage{graphicx}

\begin{document}

\title{Yet on statistical properties of traded volume: correlation and mutual information at different value magnitudes}
\author{S.~M.~Duarte~Queir\'{o}s \footnote{corresponding author}}
\email[e-mail address: ]{sdqueiro@cbpf.br}
\affiliation{Centro Brasileiro de Pesquisas F\'{\i}sicas, 150, 22290-180, Rio de Janeiro
- RJ, Brazil}
\author{Luis G. Moyano}
\email[e-mail address: ]{moyano@cbpf.br}
\affiliation{Centro Brasileiro de Pesquisas F\'{\i}sicas, 150, 22290-180, Rio de Janeiro
- RJ, Brazil}
\date{\today}

\begin{abstract}
In this article we analyse linear correlation and non-linear dependence of traded volume, $v$, of the $30$ constituents of Dow Jones 
Industrial Average at different value scales. Specifically, we have raised $v$ to some real value $\alpha $ or $\beta $, which 
introduces a bias for small ($ \alpha ,\,\beta <0$) or large ($\alpha ,\,\beta >1$) values. Our results show that small values of $v$ 
are regularly \emph{anti-correlated} with values at other scales of traded volume. This is consistent with the high liquidity of the 
$30$ equities analysed and the asymmetric form of the multi-fractal spectrum for traded volume which has supported the dynamical 
scenario presented by us.
\end{abstract}



\maketitle

\section{Introduction}

Financial market analysis has become one of the most significative examples
about application of concepts associated with physics to systems that are
usually studied by other sciences~\cite{bouchaud-potters-book}. In this
sense, ideas like \textit{scale invariance} and \textit{cooperative phenomena%
} have also found significance in systems that are not described neither by
some Hamiltonian nor some other kind of equation usually associated with
Physics (\textit{e.g.}, a master equation). Although plenty of work has been
made on the analysis and mimicry of price fluctuations, less attention has
been paid to an important observable intimately related to changes in price,
the \emph{traded volume}, $v$~\cite{karpoff}. In fact, traded volume has
been coupled to price fluctuations both on an empirical or analytical way
for some time~\cite{early}. Nonetheless, a consistent analysis of intrinsic
statistical properties of traded volume appears to be first presented in
Reference~\cite{gopi-volume}. Thereafter, it has been enlarged or revisited
by different authors~\cite{vol-anal,eisler,creta,bariloche}. In this
article, we apply a generalisation of the traditional linear
self-correlation function in order to study how small, large, and about
average (frequent) values of $v$ relate between them in time. Furthermore,
we analyse non-linear dependence using a generalised measure based on
Kulback-Leibler mutual information. Our data set is made up of 1 minute
traded volume time series, running from the $1^{st}$ July 2004 to the $%
31^{st}$ December 2004, for the $30$ equities that make the Dow Jones
Industrial Average index. Aiming to avoid the well-known intraday profile,
traded volume time series were previously treated according to a standard
procedure (see \textit{e.g.}~\cite{bariloche}).

\section{Generalised linear self-correlation function}

\label{correlation} The (normalised) correlation function, generally, 
\begin{equation}
C\left( A\left( \vec{r},t\right) ,B\left( \vec{r}^{\prime },t^{\prime
}\right) \right) =\frac{\left\langle A\left( \vec{r},t\right) \,B\left( \vec{%
r}^{\prime },t^{\prime }\right) \right\rangle -\left\langle A\left( \vec{r}%
,t\right) \right\rangle \left\langle B\left( \vec{r}^{\prime },t^{\prime
}\right) \right\rangle }{\sqrt{\left\langle A\left( \vec{r},t\right)
^{2}\right\rangle -\left\langle A\left( \vec{r},t\right) \right\rangle ^{2}}%
\sqrt{\left\langle B\left( \vec{r}^{\prime },t^{\prime }\right)
^{2}\right\rangle -\left\langle B\left( \vec{r}^{\prime },t^{\prime }\right)
\right\rangle ^{2}}},  \label{selfcor}
\end{equation}%
represents a useful analytical form to evaluate how much two random
variables depend, linearly, on each other. Leaving out spatial dependence,
when $A$ and $B$ are the same observable, Eq.~(\ref{selfcor}) represents the
straightforwardest way to appraise memory in the evolution of $A$. In any
case, it does not give us any information about the role of magnitudes.
Inspired by multi-fractal analysis~\cite{scaling}, a simple way to quantify
this type of correlation can be defined by introducing a \textit{generalised
self-correlation function}, $C\left( \hat{A}\left( t\right) ,\tilde{A}\left(
t^{\prime }\right) \right) \equiv C_{\alpha ,\beta }\left( A\right) $, where 
$\hat{A}\left( t\right) =\left\vert A\left( t\right) \right\vert ^{\alpha }$%
, $\tilde{A}\left( t\right) =\left\vert A\left( t\right) \right\vert ^{\beta
}$ (with $\alpha ,\beta \neq 0$ $\in \Re $), and $t^{\prime }=t+\tau $~%
\footnote{%
Hereon $A\left( t\right) $ is assumed to be a stationary time series. The
dependence on the \textit{waiting time}, $t$ represents an indication of
non-stationarity in the signal.}. As an example let us assume $\beta =1$.
For values of $\alpha $ greater than $1$, small values of $A$ become even
smaller and their weight in the value of $C_{\alpha ,\beta }\left( A\right) $%
, due to $ \hat{A}\left( t\right) ,\tilde{A}\left( t^{\prime
}\right) $, approaches negligibility ({\it e.g.}, when $\alpha =2$, $%
v=10^{-1}>v^{\alpha }=10^{-2}$ and $v=10<v^{\alpha }=10^{2}$). Otherwise,
when $\alpha $ is negative, we highlight values around zero (\textit{e.g.},
when $\alpha =-1$, $v=10^{-1}<v^{\alpha }=10^{1}$ and $v=10>v^{\alpha
}=10^{-1}$). In the end, after summing over all pairs $\left( \hat{A}\left(
t\right) ,\tilde{A}\left( t^{\prime }\right) \right) $, we verify that the
main contribution for $C_{\alpha ,1}\left( A\right) $ comes from large
values of $\left\vert A\left( t\right) \right\vert $ when $\alpha >1$ and
from small values of $\left\vert A\left( t\right) \right\vert $ when $\alpha
<0$. Accordingly, for $\alpha =\beta $, we estimate how values of the same
order of magnitude are related in time, when $\alpha \neq \beta $ we analyse
the relation between values with different magnitudes.

\begin{figure}[h]
\centering 
\includegraphics[width=0.4\columnwidth,angle=0]{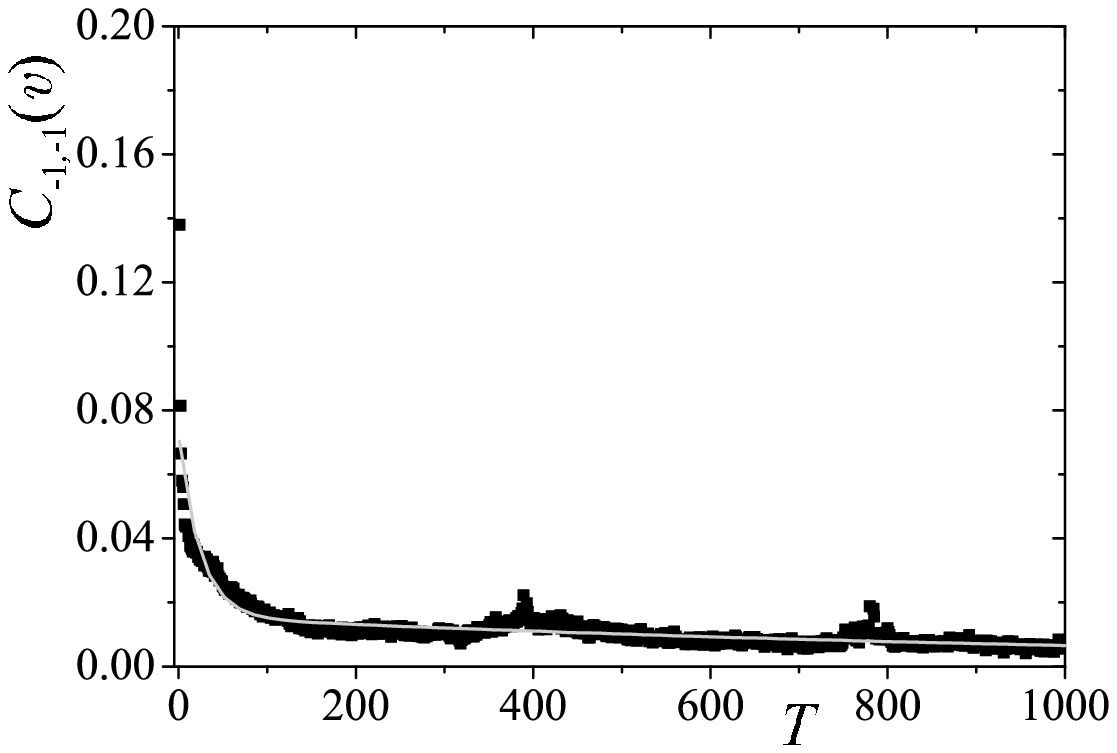} %
\includegraphics[width=0.4\columnwidth,angle=0]{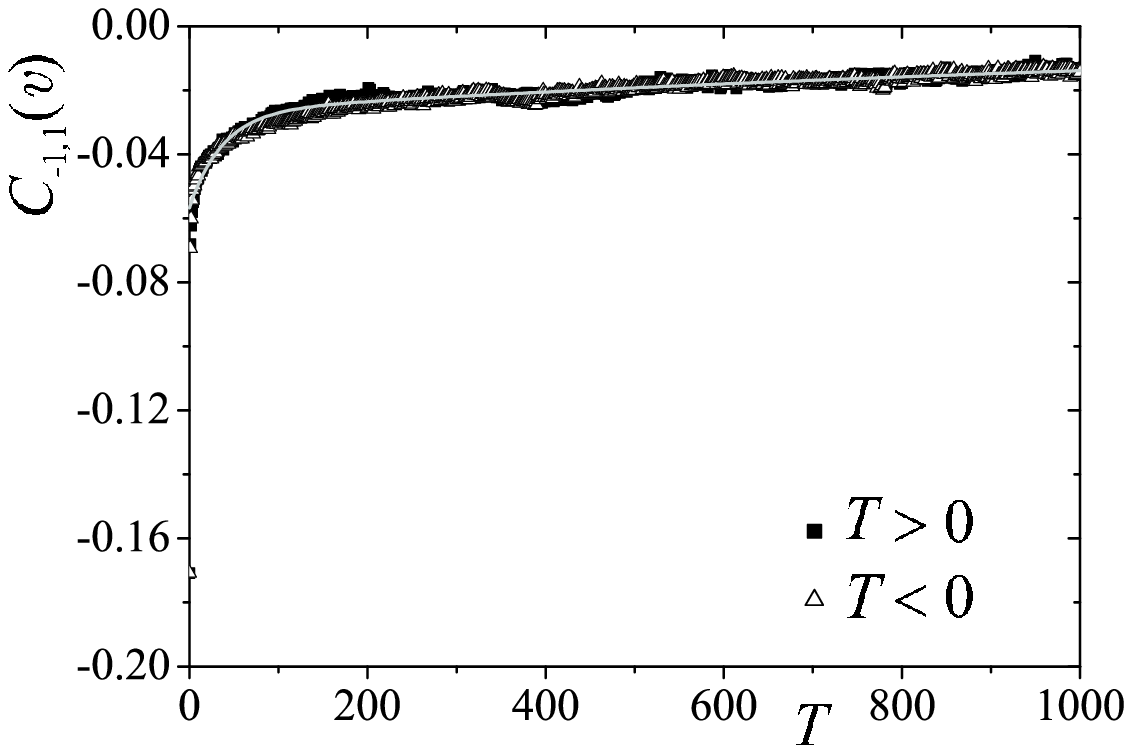} %
\includegraphics[width=0.4\columnwidth,angle=0]{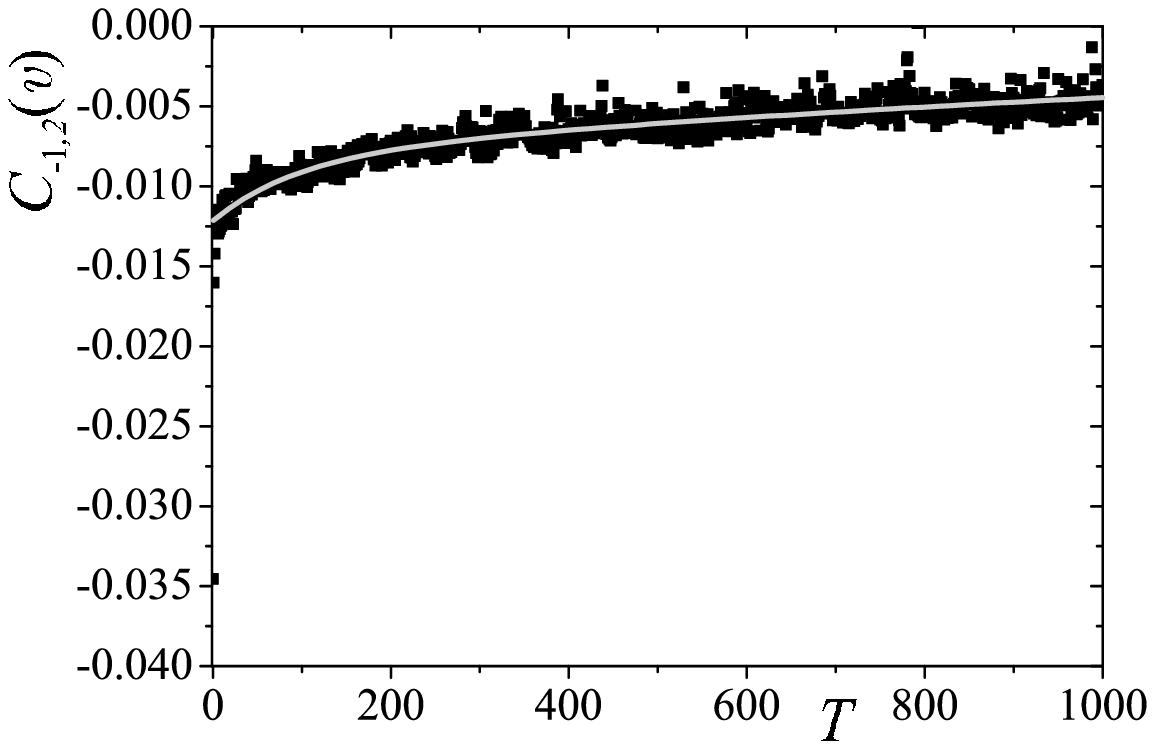} %
\includegraphics[width=0.4\columnwidth,angle=0]{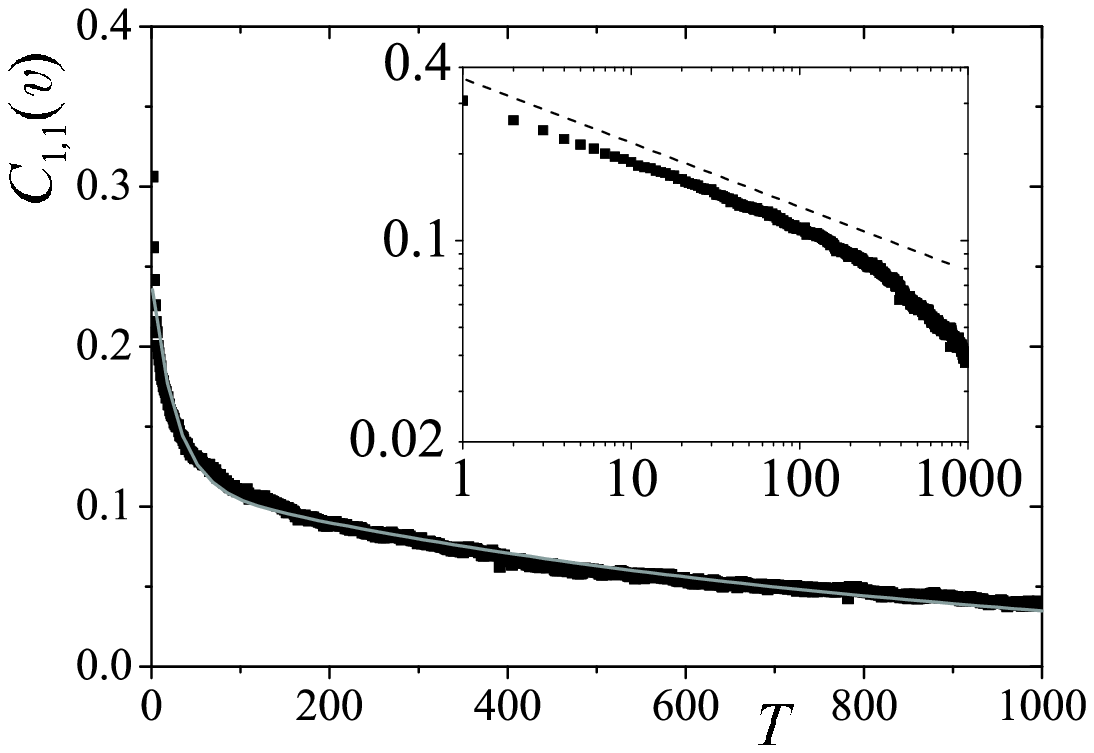} %
\includegraphics[width=0.4\columnwidth,angle=0]{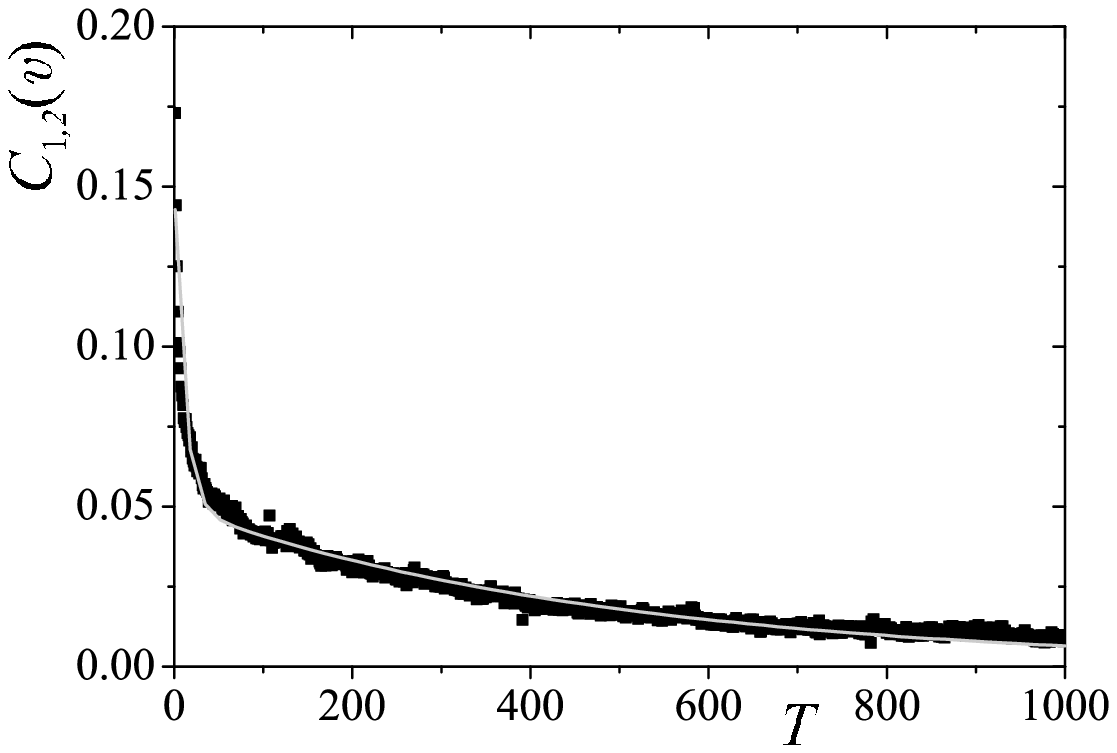} %
\includegraphics[width=0.4\columnwidth,angle=0]{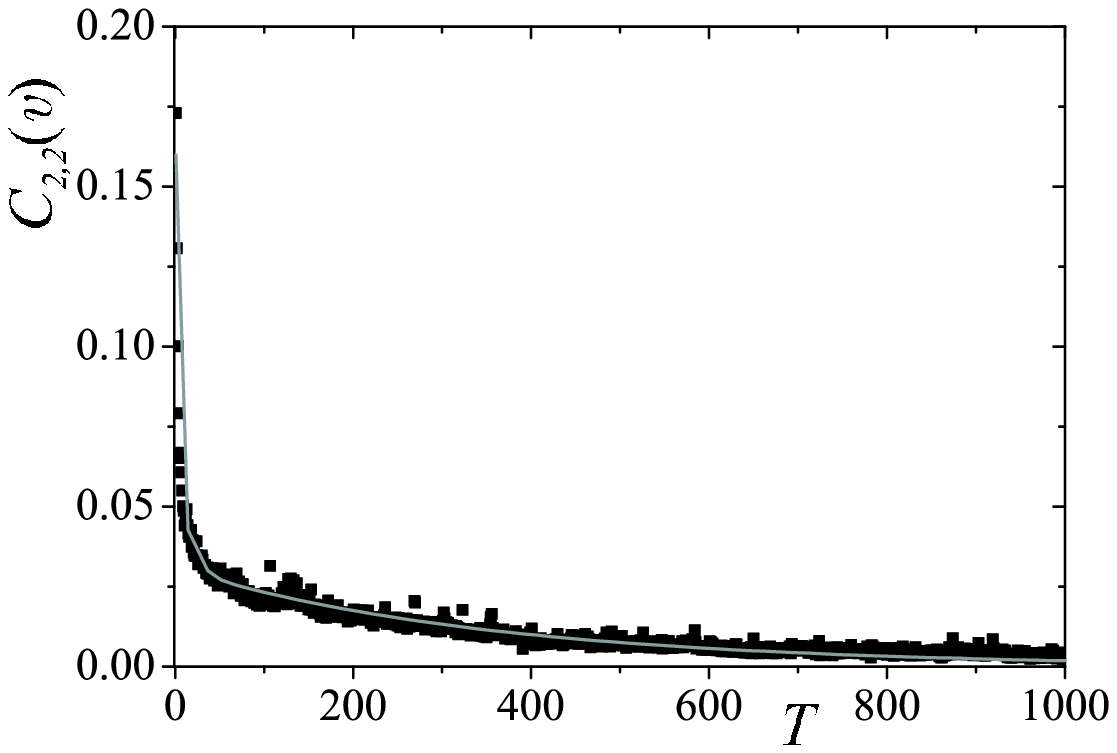}
\caption{$C_{\protect\alpha ,\protect\beta }(v)$ \textit{vs.} $T$ for the
values of $\protect\alpha $ and $\protect\beta $ presented in Table~\protect
\ref{tabela} (clockwise). On each panel black symbols are the values
obtained from time series and the grey line represents the numerical fit of $%
C_{\protect\alpha ,\protect\beta }(v)$ for a double exponential. In panel
for $C_{-1,1}(v)$ the curves for $T>0$ and $T<0$ concur, which goes along
the lines of time symmetry. The inset on $C_{1,1}(v)$ panel is a log-log
representation of the main panel. As it is visible the correlation function
does not present power-law behaviour. The same happens for all the other
values of $(\protect\alpha ,\protect\beta )$ studied.}
\label{q-product}
\end{figure}

In Fig.~\ref{q-product} we depict the results that we have obtained by
applying Eq. (\ref{selfcor}), with different pairs of $\left( \alpha ,\beta
\right) $ in traded volume time series. In Table~\ref{tabela} we present the
values of the numerical adjustment of $C_{\alpha ,\beta }\left( A\right) $
for a double exponential function, $$f\left( x\right) =a\exp \left[ -\frac{x}{%
\tau _{1}}\right] +b\exp \left[ -\frac{x}{\tau _{2}}\right] .$$ We have set
as minimum and maximum values for the exponents $-1$ and $2$. Our choice is
justified by the fact they are both able to evaluate the influence of small
and large values of $v$, and to preserve a reliable statistics.

From the analysis of figures in Table~\ref{tabela}, we observe that small
values, $\alpha (\beta )=-1$, are always anti-correlated with both frequent (%
$\alpha (\beta )=1$), and large ($\alpha (\beta )=2$) values of traded
volume. We verify that there is temporal symmetry, which can be checked if
we change $\alpha \leftrightarrow \beta $. When $\alpha $($\beta $) equal $%
-1 $, the second scale of relaxation is consistently much larger than the
observed when both exponents are positive. In addition, the values for
coefficients $a$ and $b$ (in modulus) are smaller when at least one of the
exponents is $-1$. This indicates that, besides presenting a negative
influence over frequent and large values of $v$, such an influence is
restrictable. On the other side, we have observed a very fast first decay of
the correlation function for $\beta =2$ and $\alpha =1,2$ followed by a
slower decay, though faster when compared with $\alpha =\beta =1$. This
might be interpreted as a consequence of the low frequency in large values
of $v$. This richness and disparity in behaviour for small and large values
is congruous with a previous multi-fractal analysis of $v$~\cite{bariloche}.
In this analysis, it has been observed a strong asymmetry in multi-fractal
spectrum, that has been associated with the existence of different dynamical
mechanisms prompting small and large values for trading volume~\cite%
{creta,volume}.

\begin{center}
\begin{table}[tbp]
\caption{Values of the parameters of adjust of $C_{\protect\alpha ,\protect%
\beta }\left( A\right) $ for a double exponencial, $f\left( x\right) =a\exp %
\left[ -\frac{x}{\protect\tau_{1}}\right] +b\exp \left[ -\frac{x}{\protect%
\tau _{2}}\right] $ for the results presented on the panels of Fig.~\protect
\ref{q-product}.}
\label{tabela}%
\begin{tabular}{|c|c||c|c|c|c|c|c|}
\hline
$\alpha $ & $\beta $ & $a$ & $\tau _{1}$ & $b$ & $\tau _{2}$ & $\chi ^{2}$ & 
$R^{2}$ \\ \hline\hline
$-1$ & $-1$ & $0.052\pm 0.001$ & $29\pm 4$ & $0.015\pm 0.002$ & $1138\pm 11$
& $2.0\times 10^{-5}$ & $0.78$ \\ \hline
$-1$ & $1$ & $-0.030\pm 0.001$ & $37\pm 1$ & $-0.027\pm 0.002$ & $1526\pm 22$
& $1.5\times 10^{-6}$ & $0.97$ \\ \hline
$-1$ & $2$ & $-0.004\pm 0.001$ & $94\pm 9$ & $-0.008\pm 0.001$ & $1628\pm 17$
& $5.8\times 10^{-7}$ & $0.84$ \\ \hline
$1$ & $1$ & $0.128\pm 0.002$ & $27\pm 1$ & $0.111\pm 0.001$ & $844\pm 7$ & $%
2.0\times 10^{-5}$ & $0.99$ \\ \hline
$1$ & $2$ & $0.102\pm 0.002$ & $11\pm 1$ & $0.050\pm 0.001$ & $488\pm 4$ & $%
6.5\times 10^{-6}$ & $0.97$ \\ \hline
$2$ & $2$ & $0.165\pm 0.002$ & $4\pm 1$ & $0.030\pm 0.001$ & $354\pm 5$ & $%
4.7\times 10^{-6}$ & $0.96$ \\ \hline
\end{tabular}
\end{table}
\end{center}

\section{Generalised mutual information applied to DJ30 traded volume time
series}

In information theory, the Kullback-Leibler (KL) mutual information~\cite%
{kullback-leibler-entropy} (or information gain, or information divergence)
is a distance measure (but \emph{not} a metric distance) that provides the
mean change of information related to any two probability distributions, $p$
and $p^{\prime }$. If we have, say, two experiments, with a given set of
discrete outcoming probability distributions $p$ and $p^{\prime }$,
respectively, then, the KL mutual information might be defined as,
\begin{equation} 
K(p,p^{\prime })\equiv -\sum_{j}p_{j}\ln \frac{p_{j}^{\prime }}{p_{j}}, 
\end{equation} 
where $p_{j}$ ($p_{j}^{\prime }$) is the probability of outcome $j$ in
experiment one (two). As a special case, we consider two random variables $x$
and $y$ and we set $p$ to be the joint probability distribution, $p=p(x,y)$
and $p^{\prime }$ the product of the marginal probability distributions, $%
p^{\prime }=p_{1}(x)p_{2}(y)$. In this particular case, the KL mutual
information is usually referred to as \emph{mutual information} (we will
denote it as $I(x,y)$) and it is a useful and natural tool to measure the
degree of statistical dependence between two random variables also applied
in financial analysis~\cite{tugas}.

When considering traded volume, and since we are dealing with correlated
non-linear processes, a natural way of generalising the KL mutual
information can be achieved by replacing the usual statistical theory for
the non-extensive statistical theory~\cite{tsallis}. In this generalisation,
the usual logarithm must be replaced by the $q$-logarithm, defined as $\ln
_{q}x=\frac{x^{1-q}-1}{1-q}$. When $q\rightarrow 1$, the $q$-logarithm
becomes the usual one. 
% In this way, the generalised KL mutual information~\cite{ct-kl-1998} takes 
%the form % 
%\begin{equation} 
%$I_{q}(p,p^{\prime })=-\sum_{j}p_{j}\ln _{q}\frac{p_{j}^{\prime }}{p_{j}}.$ 
%\end{equation} 
%

For $q>0$, there exist well defined minimum and maximum values of $%
I_{q}(p,p^{\prime })$ corresponding to minimum and maximum dependence
degrees between random variables, \textit{e.g.}, $x$ and $y$. This allows us
to define a criterion for statistical testing~\cite{borland-plastino-tsallis}
through the normalised quantity $R\equiv \frac{I_{q}}{I_{q}^{max}},\,\in
\lbrack 0,1]$. Its extreme value $R=0$ ($R=1$) corresponds to zero (full)
dependence between $x$ and $y$. Given $x$ and $y$, the ratio $R$ can be
calculated as a function of $q$. Typically, $R$ varies smoothly and
monotonically from 0 to 1, its two limiting values. The inflexion point in $%
R(q)$ determines the value of $q$ for which $R$ most sensibly detects
changes in the correlation between $x$ and $y$. We call this value of $q$ as 
\textit{optimal value}, $q^{op}$. It can be seen~\cite%
{borland-plastino-tsallis} that for one-to-one dependence we have $q^{op}=0$%
, and $q^{op}=\infty $ for total independence.

The generalised mutual information $R$ has already been used in~\cite{creta}
applied to traded volume time series from the components of the Dow Jones 30
index. In order to compare this quantity with the self-correlation function,
we have considered $x$ to be the time series and $y$ the same time series
with a lag in time, $\tau \equiv T$. Here, we have further analysed this
data by performing this calculation on the same lines as in section \ref%
{selfcor}, \textit{i.e.}, we have defined our random variables by modifying
the (normalised detrended) traded volume $v$ through exponents $\alpha $ and 
$\beta $, \textit{i.e.}, $x_{\alpha }=v_{j}^{\alpha }$ and $y_{\beta
}=v_{j}^{\beta }$. Then, we have computed $R$ with the same exponents as in
the section~\ref{correlation}. Our procedure can be summarised as follows:
We have first derived the probability distributions for each component time
series $i$ and its lagged counterpart with lag $\tau $. To construct the
PDFs, we have set the bin size (or, in physics terms, the coarse-graining)
to be $\Delta x=\Delta y=0.02,\forall i$. We then have calculated $R_{i}$ as
a function of the index $q$. From this, we have extracted an optimal index $%
q_{i}^{op}(\tau ,\alpha ,\beta )$ for each component $i$. Finally, we have
computed the mean $q^{op}$ value from all 30 components, \textit{i.e.} $%
q^{op}=\frac{1}{30}\sum_{i=1}^{30}q_{i}^{op}(\tau ,\alpha ,\beta )$~%
\footnote{%
Although it has been proved that statistical features depend on the
liquidity~\cite{eisler}, our averaging is completely justified since our
companies present trading values (per minute) within the same class~\cite%
{eisler-com}.}.

\begin{figure}[h!]
\centering 
\includegraphics[angle=-90,width=0.8\textwidth]{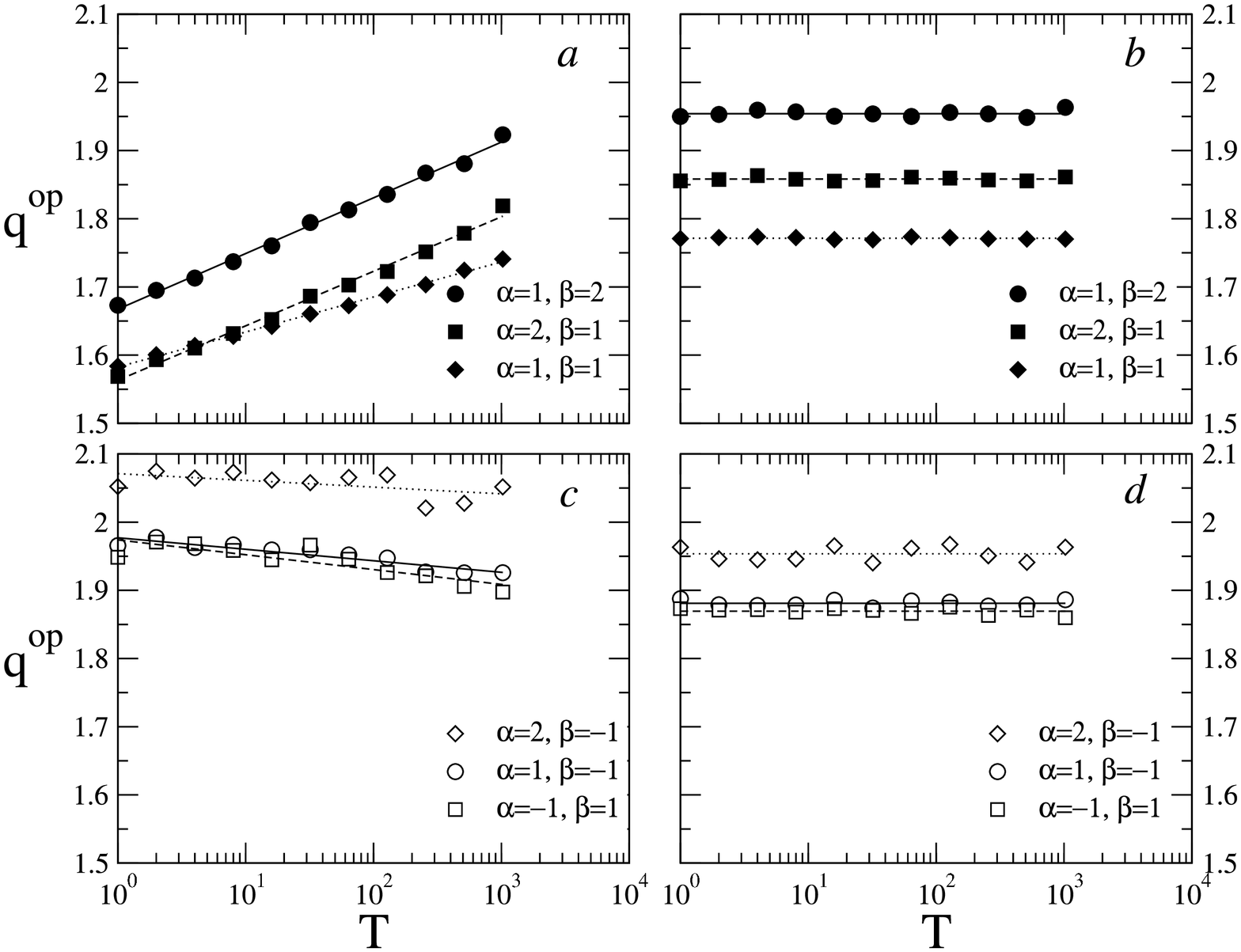}
\caption{ Optimal index $q^{op}$ versus lag $\protect\tau$.
Panel $a$: Lines correspond to fitting function $q^{op}=A+B\log\protect\tau$%
, where ($A, B$) is ($1.667\pm0.003, 0.035\pm0.001$) (solid line), ($%
1.563\pm0.004, 0.035\pm0.001$) (dashed line) and ($1.583\pm0.001,
0.0223\pm0.0003$) (dotted line) for ($\protect\alpha, \protect\beta$)=
(2,1), (1, 2) and (1, 1) respectively. Panels $b$: Same as in $a$ on the
shuffled version of the time series. Constant values are: $1.954\pm0.002$
(solid line), $1.858\pm0.001$ (dashed line) and $1.7713\pm0.0008$ (dotted
line). Panel $c$: Logarithmic fitting as in $a$: ($A, B$) is ($%
1.977\pm0.004, -0.007\pm0.001$) (solid line), ($1.974\pm0.008,
-0.009\pm0.002 $) (dashed line), ($2.07\pm0.01, -0.004\pm0.002$) (dotted
line) for ($\protect\alpha, \protect\beta$)= (2,-1), (1, -1) and (-1, 1)
respectively. Panel $d$: Same as in $c$ on the shuffled version of the time
series. Constant values are: $1.881\pm0.002$ (solid line), $1.869\pm0.002$
(dashed line) and $1.953\pm0.005$ (dotted line). }
\label{all}
\end{figure}

In Fig. \ref{all} we present our results for different values of $\alpha $
and $\beta $. Firstly we plot, for comparison purposes, the unmodified case $%
\alpha =1,\beta =1$ (panel $a$). There is a clear logarithmic dependence of $%
q^{op}$ as a function of the lag $\tau $. In the same panel we plot our
results for $\alpha =2,\beta =1$, and its symmetric case $\alpha =1,\beta =2$%
. We obtain again a logarithmic behaviour, but both additive and
multiplicative fitting parameters change. The rate of change is higher
indicating that this particular choice of $\alpha $ and $\beta $ accelerates
the loss of dependence in time. For $\alpha =1,\beta =2$, the multiplicative
parameter is very close to $\alpha =2,\beta =1$ case (see caption),
reflecting the same kind of symmetry observed in the section~\ref%
{correlation}. In panel $c$ we present results where $\alpha >0,\beta <0$ or 
$\alpha <0,\beta >0$. Our results show that, in this case, $q^{op}$
diminishes as a function of the lag, but the rate of change is not as high
as in the $\alpha >0,\beta >0$ case (see caption). Note that this result
occurs for the same exponents where anti-correlated behaviour is found
(section~\ref{correlation}) suggesting that anti-correlation might imply on
negative slope in the logarithmic behaviour. This possibility will be
verified in future work, namely on the analysis of the dependence between
volatility and traded volume~\cite{progress}.

To further analyse the meaning of these results, we have performed the same
calculations on a shuffled version of the same time series, \textit{i.e.}
applying a random reordering (in time) on each component time series. We
show our results in Fig.~\ref{all}, panels $b$ and $d$, where we have use
the same exponents as in panels $a$ and $c$ respectively. This shuffling
procedure destroys causality and in every case $q^{op}$ looses its
dependence with $\tau$. In all cases, the curves obtained from the
unshuffled data evolve towards the shuffled ones, probably reaching them for
high values of $\tau$. Thus, one can consider that $q^{op}$ obtained from
the shuffled time series act as saturation values of the unshuffled case,
when all dependence is lost.

\section{Final remarks}

%\medskip

To conclude, we have applied a generalised form of correlation function, $%
C_{\alpha , \beta}(.)$, in order to evaluate how values having different
magnitudes influence each other. The results obtained point out that small
values of traded volume are consistently anti-correlated with frequent and
large values. Moreover, frequent and large values are positively correlated.
These results are in accordance with the strong asymmetry of the
multi-fractal spectrum, which has supported our dynamical scenario for the
observable~\cite{creta,volume}.

We also have investigated the effect of modifying our data through $\alpha $
and $\beta $ on the Kullback-Leibler generalised mutual information, and its
associated optimal index $q^{op}$. Our results show that there is a
logarithmic dependence of $q^{op}$ with lag in the positive exponents case ($%
\alpha >0,\beta >0$), with different fitting parameters depending on these
exponents. In the case of negative $\alpha $ or $\beta $, we have observed
that $q^{op}$ diminishes with lag. A further analysis on this intriguing
behaviour is certainly welcome.

\medskip

We thank C.~\textsc{Tsallis} for several conversations on the subjects
treated along this manuscript. SMDQ aknowledges previous discussions about
multi-scaling with E.~M.~F. \textsc{Curado} and F.~D. \textsc{Nobre}, and Z. 
\textsc{Eisler} for has provided some of the results in manuscript of Ref.~%
\cite{eisler-com}. We also thank Olsen Data Services for the data provided
and used herein. LGM is thankful to the International Christian University
in Tokyo for the warm hospitality. Financial support from FCT/MCES
(Portuguese agency) and infrastructural support from PRONEX/CNPq (Brazilian
agency) are also acknowledged.


\begin{thebibliography}{99}
\bibitem{bouchaud-potters-book} J.-P. Bouchaud and M. Potters, \textit{%
Theory of Financial Risks: From Statistical Physics to Risk Management}
(Cambridge University Press, Cambridge, 2000); R.N. Mantegna and H.E.
Stanley, \textit{An introduction to Econophysics: Correlations and
Complexity in Finance } (Cambridge University Press, Cambrigde, 1999); J.
Voit, \textit{The Statistical Mechanics of Financial Markets}
(Springer-Verlag, Berlin, 2003)

\bibitem{karpoff} J.M. Karpoff, J. Finan. Quantitat. Anal. \textbf{22}, 109
(1997)

\bibitem{early} M.F.M. Osborne, Oper. Res. \textbf{7}, 145 (1959); C.J.W.
Granger and O. Morgenstern, Kyklos \textbf{16} (1), 1 (1963); C.C. Ying,
Econometrica \textbf{34}, 676 (1966)

\bibitem{gopi-volume} P. Gopikrishnan, V. Plerou, X. Gabaix, and H.E.
Stanley, Phys. Rev. E \textbf{62}, R4493 (2000)

\bibitem{vol-anal} R. Osorio, L. Borland and C. Tsallis in:\textit{\
non-extensive Entropy - Interdisciplinary Applications}, edited by: M.
Gell-Mann and C. Tsallis (Oxford University Press, New York, 2004)

\bibitem{eisler} Z. Eisler and J. Kert\'{e}sz, Eur. Phys. J. B \textbf{51},
145 (2006)

\bibitem{creta} J. de Souza, L.G. Moyano, and S.M. Duarte Queir\'{o}s, Eur.
Phys. J. B \textbf{50}, 165 (2006)

\bibitem{bariloche} L.G. Moyano, J. de Souza, and S.M. Duarte Queir\'{o}s,
Physica A \textbf{371}, 118 (2006)

\bibitem{scaling} J. Feder, \textit{Fractals} (Plenum, New York, 1988);
A.-L. Barab\'{a}si and T. Vicsek, Phys. Rev. A \textbf{44}, 2730 (1991); M.
Pasquini and M. Serva, Econ. Lett. \textbf{65}, 275 (1999)

\bibitem{volume} S.M. Duarte Queir\'{o}s, Europhys. Lett. \textbf{71}, 339
(2005)

\bibitem{kullback-leibler-entropy} S. Kullback and R.A. Leibler, Ann. Math.
Stat. \textbf{22}, 79 (1961)

\bibitem{tugas} C. Granger and J. Lin., J. Time Ser. Anal. \textbf{15}, 371
(1994); A. Dionisio, R. Menezes and D.A. Mendes, Physica A \textbf{344}, 326
(2004)

\bibitem{tsallis} C. Tsallis, J. Stat. Phys. \textbf{52}, 479 (1988);
Related bibliography at: \texttt{http://tsallis.cat.cbpf.br/biblio.htm}

\bibitem{ct-kl-1998} C. Tsallis, Phys. Rev. E \textbf{58}, 1442 (1998)

\bibitem{borland-plastino-tsallis} L. Borland, A.R. Plastino and C. Tsallis,
J.Math. Phys. \textbf{39}, 6490 (1998); [Erratum: J. Math. Phys. \textbf{40}%
, 2196 (1999)]

\bibitem{eisler-com} Z. Eisler, private communication on manuscript: Z.
Eisler and J. Kert\'{e}sz, \texttt{arXiv:physics/0606161} (preprint, 2006)

\bibitem{progress} S.M. Duarte Queir\'{o}s and L.G. Moyano, work in progress.
\end{thebibliography}
\end{document}